\documentclass[prb,aps,twocolumn,draft,showpacs,floatfix]{revtex4}
\usepackage{colordvi}
\usepackage{psfig}
\usepackage{amssymb}
\usepackage{amsmath}
\usepackage{colordvi}

\begin{document}
\title{Spin relaxation under identical Dresselhaus and Rashba coupling
  strengths in GaAs quantum wells}
\author{J. L. Cheng}
\affiliation{Hefei National Laboratory for Physical Sciences at
  Microscale, University of Science and Technology of China, Hefei,
  Anhui, 230026, China}
\affiliation{Department of Physics,
University of Science and Technology of China, Hefei,
  Anhui, 230026, China}
\altaffiliation{Mailing Address}
\author{M. W. Wu}
\thanks{Author to whom correspondence should be addressed}%
\email{mwwu@ustc.edu.cn.}
\affiliation{Hefei National Laboratory for Physical Sciences at
  Microscale, University of Science and Technology of China, Hefei,
  Anhui, 230026, China}
\affiliation{Department of Physics,
University of Science and Technology of China, Hefei,
  Anhui, 230026, China}
\altaffiliation{Mailing Address}

\date{\today}
\begin{abstract}
Spin relaxation under identical Dresselhaus and Rashba
  coupling strengths in GaAs quantum wells is studied
in both the traditional collinear statistics, where the energy spectra
do not contain the spin-orbit coupling terms, and the helix statistics, where
the spin-orbit couplings are included in the energy spectra.
We show that there is only marginal difference between
the spin relaxation times obtained from these two different statistics.
We further show that with the cubic term of the Dresselhaus spin-orbit coupling
included, the spin relaxation time along the (1,1,0) direction becomes finite,
although it is still much longer than that along the other two perpendicular
directions. The properties of the spin relaxation along this special direction
under varies conditions are studied in detail.

\end{abstract}
\pacs{72.25.Rb,72.15.Lh,71.10.-w,67.57.Lm}

\maketitle
\section{Introduction}
Semiconductor spintronics which aims at
making use of electron spin degrees of freedom has attracted much
attention both theoretically and experimentally\cite{spintronics,das} owing to its
potential applications such as spin transistors.\cite{Datta1990,Loss}  The
realization of such device requires a long spin relaxation/dephasing (R/D) time which
should be at least longer than the time of one expected operation.
Therefore it is essential to understand the mechanism of the
spin R/D and to find effective ways to manipulate the spin relaxation
time (SRT) and/or spin dephasing time.

It is well known that in $n$-type Zinc-blende semiconductors,
the leading spin R/D mechanism is the so called
D'yakonov-Perel' (DP) mechanism,\cite{dp} which is due to
the spin state splitting of the conduction band at $k\not=0$ from
the spin-orbit interaction in crystals without inversion center.
This splitting is equivalent to an effective magnetic field (EMF) ${\bf h}({\bf
k})$ acting on the spin,
with its magnitude and orientation depending on ${\bf k}$. In quantum wells (QW's)
this EMF is composed of the Rashba term\cite{rashba} due to
the lack of the structure inversion symmetry and the
Dresselhaus term\cite{dress} due to the lack of the bulk inversion symmetry.
Their contributions to the energy spectra\cite{winkler,averkiev}
 as well as the spin R/D have been widely studied in the
literature.\cite{averkiev,meier,wu1,wu,lau,ivan,frank,wugo,wu3,wu2,wu21,frank1,kri,ohms,ob}
Under some special conditions, such as when the growth direction
of the QW is along  (110) axis\cite{dohrmann,Cartoixa,wugo} or when the growth
direction is along (001) axis but the strengths
of the Rashba and the Dresselhauls terms are
comparable,\cite{Loss,averkiev,winkler,averkiev1999,flatte,Ehud,Loss1,Loss2} the SRT
shows strong anisotropy and along some special polarization
direction one can get extremely long SRT (It can be infinite when the cubic terms
in the Dresselhaus term is ignored and the strengths of the  Rashba and the
Dresselhauls terms are identical).
Schliemann {\em et al.} further
proposed a new spin transistor based on this strong anisotropy of the SRT when
the strengths of the Rashba and the Dresselhauls terms are identical.\cite{Loss}
Very recently Jiang and Wu proposed to use strain to get extremely long spin dephasing time
(also relaxation time) in GaAs (001) QW's where there is no anisotropy along
different directions.\cite{strain}

Among these studies, Wu proposed a many-body approach by setting up the kinetic spin Bloch
equations\cite{wu1,wu} and solving it self-consistently with all the scattering,
{\em i.e.}, the electron-electron, electron-phonon and electron-nonmagnetic impurity
scattering explicitly included. By this approach, one not only gets the spin R/D time
due to the single-particle effective spin-flip scattering, but also obtains
that due to the many-body effects\cite{wumany,wu2,wu21} which have shown to be dominant
in some $n$-type semiconductor QW's\cite{wu2,wu21,wumany1} and give many different
behaviors from the single-particle
spin R/D both in spin precession\cite{wu2,wu21,wumany1} and spin transport.\cite{wu3}
The many-body effect here is not only referred to be the effect of the Coulomb scattering,
 but also expanded to include the spin R/D due to
 the inhomogeneous broadening from  the EMF\cite{wumany,wu2,wu21}
combined with the spin conserving scattering.
Moreover, this approach also includes the counter-effect from the scattering to the
inhomogeneous broadening from the anisotropic EMF, which has
been shown to be also very important in the spin kinetics.\cite{wu2,wu21}
In this paper, we apply this approach to study the spin relaxations in
GaAs (001) QW's with the identical Rashba and Dresselhaus spin-orbit coupling strengths,
especially at the case when the cubic term of the Dresselhaus term is included. The
later has been investigated in the single-particle picture\cite{Loss1,Loss2,averkiev}
without counting the many-body effects.

Another issue we address in this paper is the influence of the
two spin eigenstates on the
spin kinetics. It is noted that in our previous
works\cite{wu1,wu,wu3,wu2,wu21,wumany,wumany1,strain}
 as well as many other works\cite{averkiev,meier,lau,frank,frank1,kri}
 in the literature, the equilibrium state
is taken as the Fermi distribution of electrons in conduction band
without the spin-orbit coupling from the DP term. Therefore, the
energy spectrum of electrons is $\varepsilon_k=k^2/2m^\ast$,
with $m^\ast$ denoting the effective mass of electrons, and
the eigenstates of spin are the eigenstates of
$\sigma_z$, {\em i.e.} $\chi_\uparrow=(1,0)^T$
and $\chi_\downarrow=(0,1)^T$,
which are collinear in the laboratory coordinates.
In the following we refer the Fermi distribution composed by
these eigenstates (without the spin-orbit coupling) to be the
collinear Fermi distribution and the statistics to be the
collinear statistics.
In the meantime, there is another equilibrium state
used in the literature  which is  the
Fermi distribution of electrons with the eigenstates being those of
electrons in the conduction band with
the DP spin-orbit coupling.
Therefore the energy spectrum is now being
\begin{equation}
\label{spectrum}
\varepsilon_{{\bf k},\eta}=
k^2/2m^\ast+\eta|{\bf h}({\bf k})|
\end{equation}
 with $\eta=\pm1$ for the two spin branches.
The eigenfunctions of spin are now
\begin{equation}
\label{helix}
|\eta\rangle=\frac{1}{\sqrt{2}}
[\chi_{\uparrow} + \eta\frac{{\tilde h}(\mathbf{k})}
{|{\bf h}(\mathbf{k})|}\chi_{\downarrow}]
\end{equation}
with ${\tilde h}({\bf k})
=h_x({\bf k})+ih_y({\bf k})$.
The spin polarizations therefore strongly depend on
the momentum ${\bf k}$. In the following
we refer the Fermi distribution composed by
these eigenstates (with the spin-orbit coupling)
to be the helix Fermi distribution and the statistics to be the
 helix statistics as the eigenstates shows helix spin
structure when the DP EMF is composed only by the Rashba term.\cite{cheng}
Some works are performed with the helix statistics especially
 those dealing with spin
relaxation in quantum dots\cite{cheng1,lu} and
spin Hall conductivity.\cite{zhang,liu,Sinitsyn} Very recently
Lechner and R\"{o}ssler\cite{lechner} and Grimaldi\cite{Grimaldi} gave some
formal theory to discuss the
spin R/D using the helix statistics.
However, whether the collinear statistics widely used in the spin
R/D calculation is adequate
in two dimensional electron gas (2DEG)
is unknown yet. In principal, when the DP spin-orbit
coupling is weak, the collinear statistics
is good enough for the spin dephasing
calculation. However, if one comes to the strong coupling
regime which can be obtained in 2DEG
confined in narrow quantum wells,\cite{Lommer,Jusserand,Grundler,Sato} then two
things happen: One is that the strong DP EMF causes strong interference effect
and induces strong spin depolarization which is irrelevant  to
the spin dephasing effect; Another
is that the statistical equilibrium state might have to be changed into
that in the helix statistics. The later then changes all the
scattering and {\em might} change
the spin R/D when the DP EMF is strong.
Nevertheless, it is not so obvious as
when the EMF is very strong, the spin polarization decreases
dramatically due to
the interference effect long before the scattering plays its role. In this
paper we address this issue by solving the
kinetic spin Bloch equations with both collinear and helix statistics.

We organize the paper as follows: We present our model and kinetic spin
Bloch equations with both the helix and the collinear statistics in Sec.\ II.
Then we present our numerical results in Sec.\ III. We first
compare the spin relaxation with two spin statistics in Sec.\ \ref{sectiona}
and \ref{sectionb}, but
without the Coulomb scattering.
Then we discuss the spin relaxation under the identical Dresselhaus
and Rashba coupling  strengths in GaAs QW's with the collinear statistics
by including all the scattering.
We conclude in Sec.\ IV.

\section{Model and kinetic spin Bloch equations}
We construct the kinetic spin Bloch equations\cite{wu1,wu} by using
the nonequilibrium Green function method\cite{haug} with these
  two different statistics and write them in the unified form as follows:
\begin{equation}
  \label{eq:Bloch}
  \dot{\rho}_{\mathbf{k}}
  + \dot{\rho}_{\mathbf{k}}|_{\mathtt{coh}}
  +\dot{\rho}_{\mathbf{k}}|_{\mathtt{scatt}}\ = 0\ .
\end{equation}
Here $\rho_{{\mathbf k}}$ represents a single
particle density matrix of electrons with wavevector ${\bf k}$.
Often we project the density matrix in the laboratory
spin space (the collinear
spin space) which is
constructed by basis with spin states oriented relative to a fixed
direction, {\em i.e.},  $\chi_{\uparrow}$ and $\chi_{\downarrow}$. In
such spin space, the diagonal matrix elements
$\rho_{{\mathbf k}, \sigma,\sigma}=f_{{\mathbf k},\sigma}$
with $\sigma=\uparrow$
or $\downarrow$ describe the distribution
functions of electrons with momentum $\mathbf{k}$ and spin $\sigma$
orientating along the laboratory $z$-axis.
The off-diagonal elements $\rho_{\uparrow,\downarrow}$ describe the corrections
(coherence) between the spin-up and spin-down
states, with its real and imaginary parts standing for the spin polarizations
along the $x$- and $y$-axes respectively.

The coherent part of the two sets of Bloch equations with different
statistics are the same and are given in the collinear spin space as
\begin{equation}
\dot{\rho}_{\mathbf{k}}|_{\mathtt{coh}} =
i[H_{R}+H_{D}-\sum_{\mathbf{q}}V_{\mathbf{q}}\rho_{{\mathbf{k}} -
    {\mathbf{q}}},\ \rho_{\mathbf{k}}]\ ,
\label{eq:coh}
\end{equation}
with $H_{R(D)}={\bf h}_{R(D)}\cdot\mbox{\boldmath$\sigma$\unboldmath}$ representing
the Rashba\cite{rashba} (Dresselhaus\cite{dress}) Hamiltonian and
\boldmath$\sigma$\unboldmath standing for the
Pauli matrices.  It describes the spin precession along the direction of the
EMF.  It is noted that  the Hartree-Fock
term $\sum_{\mathbf{q}}V_{\mathbf{q}}\rho_{{\mathbf{k}} -
    {\mathbf{q}}}$ in the coherent part  cannot flip the total electron spin as
\begin{equation}
\label{sum1}
\sum_{\mathbf k}\Big[\sum_{\mathbf
  q}V_{\mathbf{q}}\rho_{\mathbf{k-q}},\rho_{\mathbf k}\Big]=0\ .
\end{equation}
Therefore the DP term is the only one that flips the total spin
polarization. For GaAs (001) QW with well width $a$ and under the
infinite-well-depth assumption,
$\mathbf{h}_{R}(\mathbf{k})=\alpha(k_y,-k_x,0)$ and
$\mathbf{h}_{D}(\mathbf{k})=\gamma(k_x[k_y^2-(\frac{\pi}{a})^2],
k_y[(\frac{\pi}{a})^2-k_x^2],0)$. Here $\alpha$ is the strength of
the Rashba term and $\gamma$ is the material-specific strength of
the Dresselhaus term.  $\alpha$ can be tuned by the external gate
voltage.\cite{rashba} By changing the external gate voltage, one
may have  $\beta=\gamma(\frac{\pi}{a})^2$.\cite{Loss,Ehud} If one
further ignores the cubic term in ${\bf h}_D({\bf k})$, at the
special polarization direction (1,1,0), Eq.\ (\ref{helix}) becomes
$|\eta\rangle=(1,\eta)^T$ which is independent on ${\bf k}$.
$V_{\mathbf{q}}$ in Eq.\ (\ref{eq:coh}) reads $V_{\bf
q}=\sum_{q_z}\frac{4\pi
e^2}{\kappa_0({\mathbf{q}}^2+q_z^2+\kappa^2)}|I(iq_z)|^2$. Here
$\kappa_0$ denotes the static electric constant and $\kappa^2=4\pi
e^2 N_e/(a k_B T)$ stands for the Debye-H\"ucke screening
constant. $N_e$ represents  the 2D electron density. The form factor
$|I(iq_z)|^2=\pi^4\sin^2y/[y^2(y^2-\pi^2)^2]$ with $y=q_za/2$. The
square bracket $[A, B]=AB-BA$ is the commutator. It is noted that
$\dot{\rho}_{\mathbf{k}}|_{\mathtt{coh}}$ is independent on the
specific spin statistics and can be expanded as
\begin{eqnarray}
\dot{f}_{\mathbf{k},\uparrow}|_{\mathtt{coh}} &=&
-\dot{f}_{\mathbf{k},\downarrow}|_{\mathtt{coh}} =
2\mbox{Im}[P^{\ast}_{\mathbf{k}}\rho_{\mathbf{k},\uparrow\downarrow}]\ ,
\\
\dot{\rho}_{\mathbf{k},\uparrow\downarrow}|_{\mathtt{coh}}
&=& -i
P_{\mathbf{k}}(f_{\mathbf{k},\uparrow}-f_{\mathbf{k},\downarrow})
- i \Delta_{\mathbf{k}}\rho_{\mathbf{k},\uparrow\downarrow}\ ,
\end{eqnarray}
with $P_{\mathbf{k}}=({\tilde
  h}^{*}(\mathbf{k})-\sum_{\mathbf{q}}V_{\mathbf{q}}\rho_{\mathbf{k-q},
  \uparrow\downarrow})$ and $\Delta_{\mathbf{k}}=\sum_{\mathbf{q}}V_{\mathbf{q}}(f_{\mathbf{k-q},\uparrow}-f_{\mathbf{k-q},\downarrow})$.

The scattering terms include the contributions from the
electron-non-magnetic-impurity scattering, the electron-phonon scattering
and the electron-electron Coulomb scattering:
\begin{widetext}
\begin{eqnarray}
  \dot{\rho}_{\mathbf{k}}|_{\mathtt{scatt}} &=&\bigg\{
  \pi N_i\sum_{\mathbf{q}}|U_{\mathbf{q}}|^2
  \sum_{\xi_1,\xi_2}
  \delta(\varepsilon_{\mathbf{k}-\mathbf{q},\xi_1}-\varepsilon_{\mathbf{k},\xi_2})
  T_{\mathbf{k}-\mathbf{q},
    \xi_1}(\rho_{\mathbf{k}}-\rho_{\mathbf{k}-\mathbf{q}})T_{\mathbf{k},\xi_2} \nonumber
\\  &+&
  \pi\sum_{\mathbf{q}q_z\lambda}|g_{\mathbf{q}q_z\lambda}|^2
  \sum_{\xi_1,\xi_2}T_{\mathbf{k}-\mathbf{q},\xi_1}
  \big\{\delta(\varepsilon_{\mathbf{k}-\mathbf{q},\xi_1}-\varepsilon_{\mathbf{k},\xi_2}
  + \Omega_{\mathbf{q}q_z\lambda})
  [(N_{\mathbf{q}q_z\lambda}+1)(1-\rho_{\mathbf{k}-\mathbf{q}})\rho_{\mathbf{k}}\nonumber\\
&&\mbox{}-
N_{\mathbf{q}q_z\lambda}\rho_{\mathbf{k}-\mathbf{q}}(1-\rho_{\mathbf{k}})]\nonumber\\
  &&\mbox{}+\ \delta(\varepsilon_{\mathbf{k}-\mathbf{q},\xi_1}
  -\varepsilon_{\mathbf{k},\xi_2}-
  \Omega_{\mathbf{q}q_z\lambda})[N_{\mathbf{q}q_z\lambda}(1-\rho_{\mathbf{k}-\mathbf{q}})
  \rho_{\mathbf{k}}-(N_{\mathbf{q}q_z\lambda}+1)\rho_{\mathbf{k}-\mathbf{q}}
(1-\rho_{\mathbf{k}})]\big\}T_{\mathbf{k},\xi_2}\nonumber
\\
 &+&
  \pi\sum_{\mathbf{q}\mathbf{k}^{\prime}}V_{\mathbf{q}}^2\sum_{\xi_1, \xi_2, \xi_3,
    \xi_4}\delta(\varepsilon_{\mathbf{k}^{\prime},\xi_3}
-\varepsilon_{\mathbf{k}^{\prime}-\mathbf{q},\xi_4}+\varepsilon_{\mathbf{k}
-\mathbf{q},\xi_1}-\varepsilon_{\mathbf{k},\xi_2})
T_{\mathbf{k}-\mathbf{q},\xi_1}\nonumber
\\
 &&\times\big\{\mbox{Tr}[T_{\mathbf{k}^{\prime},\xi_3}(1-\rho_{\mathbf{k}^{\prime}})
 \rho_{\mathbf{k}^{\prime}-\mathbf{q}}T_{\mathbf{k}^{\prime}-\mathbf{q},\xi_4}]
 (1-\rho_{\mathbf{k}-\mathbf{q}})\rho_{\mathbf{k}}\nonumber\\
&&\mbox{}
  -
  \mbox{Tr}[T_{\mathbf{k}^{\prime},\xi_3}\rho_{\mathbf{k}^{\prime}}
(1-\rho_{\mathbf{k}^{\prime}-\mathbf{q}})T_{\mathbf{k}^{\prime}-\mathbf{q},\xi_4}]
\rho_{\mathbf{k}-\mathbf{q}}(1-\rho_{\mathbf{k}})\big\}T_{\mathbf{k},\xi_2}\bigg\}
\nonumber
\\
 &+&\Big\{\cdots \Big\}^{\dagger}\ ,
 \label{scat}
\end{eqnarray}
\end{widetext}
in which $\{\cdots\}^{\dagger}$ is the Hermite conjugate
of the same terms in the previous $\{\}$.
The subscript $\xi$ denotes
the spin branch which is $\sigma=\uparrow,\downarrow$ in the
collinear statistics and $\eta=\pm$ in the helix statistics.
  $N_i$ in Eq.\ (\ref{scat}) is the
impurity density and  $|U_{\mathbf{q}}|^2=\sum_{q_z}\bigl\{4\pi Z_i
e^2/[\kappa_0 (q^2+q_z^2)]\bigr\}^2 |I(iq_z)|^2$ is the
impurity potential with $Z_i$ denoting the charge
number of the impurity.
$|g_{\mathbf{q}q_z\lambda}|^2$ and $N_{\mathbf{q}q_z\lambda}=
[\exp(\Omega_{\mathbf{q}q_z\lambda}/k_BT)-1]^{-1}$ are the  matrix element
of the electron-phonon interaction
and the Bose distribution function with phonon energy spectrum
$\Omega_{\mathbf{q}q_z\lambda}$ at phonon mode $\lambda$ and
wavevector $(\mathbf{q},q_z)$ respectively.
It is noted that the scattering
terms are different for different statistics.
With the collinear statistics,
the matrix $T_{\mathbf{k},\xi}=T_{\mathbf{k},\sigma}$
is $1/2$, a constant number, and the
energy spectrum $\varepsilon_{{\mathbf
    k},\xi}=\varepsilon_{\mathbf{k},\eta}\equiv\varepsilon_{k}$.
Then the scattering terms are exactly the same as those in Refs.\ \onlinecite{wu2,wu21}.
However,  with the helix statistics,
the scattering terms become more complicated.
$T_{\mathbf{k},\xi}=T_{\mathbf{k},\eta}$ is a $2\times2$
  matrix and becomes ${\mathbf{k}}$-dependent:
\begin{equation}
T_{{\bf k},\eta}=\frac{1}{2}[1+\eta \frac{\mathbf{h}(\mathbf{k})}
{|\mathbf{h}(\mathbf{k})|}
\cdot\mbox{\boldmath$\sigma$\unboldmath}]\ .
\end{equation}
Moreover, the energy spectrum in the $\delta$-functions in Eq.\ (\ref{scat})
$\varepsilon_{{\bf k},\eta}$ is anisotropic in ${\bf k}$
[Eq.\ (\ref{spectrum})].
When the strength of the DP term tends to
zero, the scattering terms in these
two statistics become identical.
%\Red{For the scattering term, the
%  previous works are only give the electron impurity
%  scattering\cite{Grimaldi} or the electron-phonon
%  scattering\cite{ivch} by the reduced density matrix method, but here
%  we give all the scattering term including the electron-electron scattering.}

Besides the collinear spin space, there is another spin space (the helix
spin space) used in the literature\cite{cheng,lechner} which is
 constructed by spin states
$|+\rangle$ and $|-\rangle$. The density matrix and the two sets of kinetic spin
Bloch equations with two different statistics can also be written in
the helix spin space and the physics itself is the same as that in the
collinear spin space. They are given in Appendix {\ref{helix_formula}}.

It is noted that
\begin{equation}
\label{sum2}
\sum_{\bf k}\dot{\rho}_{\mathbf{k}}\Big|_{\mathtt{scatt}}\equiv 0
\end{equation}
for both statistics as all the scattering here
is the spin-conserving scattering.

The scattering terms tend to drive the out-of-equilibrium system back to the
equilibrium one $\rho_0({\bf k}) = \{\exp([H_0({\bf k})-\mu]/k_BT)+1\}^{-1}$
with $H_0(k)=\frac{k^2}{2m^{\ast}}$  for the collinear
statistics and $H_0({\bf k})=\frac{k^2}{2m^{\ast}} +
\mathbf{h}(\mathbf{k})\cdot\mbox{\boldmath$\sigma$\unboldmath}$
for the helix statistics. This can be seen from the fact
that $\dot{\rho}_{\mathbf{k}}|_{\mathtt{scatt}}=0$ when $\rho_{\mathbf
k}=\rho_0(\mathbf k)$ for each corresponding statistics.
The equilibrium state $\rho_0$ can be further written as
\begin{eqnarray}
\label{rho0}
  \rho_0(\mathbf{k})&=&
[\exp((H_0-\mu)/k_BT)+1]^{-1}\nonumber\\ &=&
  \sum_{\xi}f(\varepsilon_{\mathbf{k},\xi}-\mu)T_{\mathbf{k},\xi}
\end{eqnarray}
for both statistics,
with $f(x)=[\exp(x/k_BT) + 1]^{-1}$ the Fermi distribution and $\mu$, the
chemical potential. For the collinear statistics, Eq.\ (\ref{rho0}) returns
to the familiar case $\rho_0({\bf k})=f(\varepsilon_{k}-\mu)$.

It is further noted that besides the energy spectrum $\varepsilon_{\mathbf{k}, \xi}$ in the two
statistics is different, the spin polarization in the equilibrium state
 ${\mathbf S}_{\mathbf
  k}^0=\mbox{Tr}[\rho_0({\mathbf k})\mbox{\boldmath$\sigma$\unboldmath}]/2$ for each
$\mathbf k$ also differs in the two statistics. For the equilibrium
states in the collinear statistics, there is not any polarization ${\mathbf S}_{\mathbf k}^0=0$
for each $\mathbf k$. However, in the helix statistics ${\mathbf S}_{\mathbf
  k}^0=\frac{\mathbf{h}(\mathbf{k})}{|\mathbf{h}(\mathbf{k})|}
[f(\varepsilon_{\mathbf{k},+}-\mu) -
f(\varepsilon_{\mathbf{k},-}-\mu)]$ gives the spin polarization along the
direction of the EMF of the DP term.
But the total spin $\mathbf{S}^0=\sum_{\mathbf k}\mathbf{S}_{\mathbf k}^0$ is still zero.

Although in different
statistics, the way of the spin relaxation {\em might be} different due to the different
equilibrium states. Here we first show analytically that
 the conclusion that when the strengths
of the Rashba and the Dresselhaus spin-orbit couplings are identical and the cubic
term of the Dresselhaus term is ignored, the SRT along the
direction $\mathbf{\hat{n}}_1=\frac{1}{\sqrt{2}}(1,1,0)$ is
infinite,\cite{Loss} does not depend on the statistics.
This is because with the identical strengths of the Rashba and the Dresselhaus
spin-orbit couplings and in the absence of the cubic term of the
Dresselhaus term, the EMF is now proportional to $\mathbf{\hat{n}}_1$:
$\mathbf{h}(\mathbf{k})=|\mathbf{h}(\mathbf{k})| \mathbf{\hat{n}}_1$
and hence $H_{R}+H_{D}=|\mathbf{h}(\mathbf{k})|
\mathbf{\hat{n}}_1\cdot\mbox{\boldmath$\sigma$\unboldmath}$. The total spin
polarization  along ${\bf \hat n}_1$ in the collinear space is $S_{\mathbf{\hat{n}}_1}(t)
={\mathbf S}\cdot\mathbf{\hat{n}}_1$, with $\mathbf{S}=\sum_{\mathbf k}\mathbf{S}_{\mathbf k}$
and ${\mathbf S}_{\mathbf
  k}=\mbox{Tr}[\rho_{\mathbf k}\mbox{\boldmath$\sigma$\unboldmath}]/2$. Then from
  Eq.\ (\ref{eq:Bloch}) one has, in both statistics,
\begin{equation}
\frac{\partial}{\partial t}S_{\mathbf{\hat{n}}_1}+i\sum_{\mathbf
  k}\mbox{Tr}([H_R+H_D, \rho_{\mathbf
  k}]\mathbf{\hat{n}}_1\cdot\mbox{\boldmath$\sigma$\unboldmath})=0\ ,
\end{equation}
with the help of Eqs.\ (\ref{sum1}) and (\ref{sum2}). It is easy to verify
that the second term of the equation is zero as $\mbox{Tr}([H_R+H_D,
\rho_{\mathbf
  k}]\mathbf{\hat{n}}_1\cdot\mbox{\boldmath$\sigma$\unboldmath})
=
|\mathbf{h}(\mathbf{k})|\mbox{Tr}[\mathbf{\hat{n}}_1\cdot\mbox{\boldmath$\sigma$\unboldmath}\rho_{\mathbf
  k}\mathbf{\hat{n}}_1\cdot\mbox{\boldmath$\sigma$\unboldmath}
- \rho_{\mathbf
  k}\mathbf{\hat{n}}_1\cdot\mbox{\boldmath$\sigma$\unboldmath}\mathbf{\hat{n}}_1
\cdot\mbox{\boldmath$\sigma$\unboldmath}]\equiv 0$.
Then the polarization
$S_{\mathbf{\hat{n}}_1}(t)=S_{\mathbf{\hat{n}}_1}(t=0)$ does not relax
in both statistics.

\section{Numerical results}

It is noted that when the cubic term of the Dresselhaus term is
included, the electron spin polarization along any direction  relaxes, even
when the strengths of the Rashba and Dresselhaus terms are identical. Moreover,
even when the spin lifetime along ${\bf \hat{n}}_1$ is infinity, the spin lifetimes
along the two perpendicular directions $\mathbf{\hat{n}}_2=\frac{1}{\sqrt{2}}(1,-1,0)$ and
$\mathbf{\hat{z}}=(0,0,1)$ are finite. It is not clear yet that how the different statistics
change these SRT's. This is what we are going to address first in following.

It is seen that all the unknowns to be solved
appear nonlinearly in the coherent and the scattering parts.
Therefore the kinetic spin Bloch equations have to be solved
self-consistently to obtain
the temporal evolutions of the electron distribution functions
$f_{\mathbf{k},\sigma}(t)$ and the spin coherence
$\rho_{\mathbf{k},\uparrow\downarrow}(t)$.
The SRT along any direction $\hat{\mathbf{n}}$ is
obtained by fitting the envelope of the quantity
$S_{\mathbf{\hat{n}}}$ under a small initial
polarization along $\mathbf{\hat{n}}$ direction
\begin{eqnarray}
\label{eq:initial_rho}
\rho_{\mathbf{k}}(t=0) &=&
\frac{1}{2}\Big[\sum_{\xi=\pm1}f(\varepsilon_{k}-\mu_{\xi})
  \nonumber\\ && +\sum_{\xi=\pm1}\xi f(\varepsilon_{k}-\mu_{\xi})
\mathbf{\hat{n}}\cdot\mbox{\boldmath$\sigma$\unboldmath}\Big]\ .
%\left(\begin{array}{cc}a&a\\a&a\end{array}\right)\ .
\end{eqnarray}
By choosing spin dependent chemical potentials $\mu_{\pm1}$, one can
construct an initial state with the given electron density
$N_e=\sum_{\mathbf{k}}\mbox{Tr}(\rho_{\mathbf{k}})$ at
$4\times10^{11}$\ cm$^{-2}$ and the initial spin  polarization
$P_{\mathbf{\hat{n}}}=S_{\mathbf{\hat{n}}}/N_e$ along the direction
$\mathbf{\hat{n}}$ at 5\ \%.

In the numerical calculation, how to compute the scattering terms are the
central problem. For the collinear statistics, this problem has been solved
successfully with high accuracy and fast CPU speed. The details of the
numerical scheme has been laid out in Ref.\ \onlinecite{wu21}.
For the helix statistics, however, things are much more complicated. This is
because in the helix energy spectrum, it is impossible to divide the
momentum space into a set of discrete grids in the way that after performing the
energy conservation ($\delta$-functions) in the scattering terms, all the
final states still belong to this set.
Therefore we still divide the momentum space by the mesh as that in
Ref.\ \onlinecite{wu21} but approximate the $\delta$-function by the Gaussian distribution
$\delta(x)=\mathop{\mathrm{lim}}\limits_{\sigma\to
  0}{\exp{(-x^2/\sigma^2)}}/(\sqrt{\pi}\sigma)$ and take the width
to be $\sigma=0.5\Delta E$ with $\Delta E$ representing the
energy interval of the grids. This approximation requires
much more grid points to converge the results. Moreover, it also
costs one more integration as the $\delta$-function cannot be carried out analytically
in the discrete space. Therefore the CPU time becomes much longer in the helix statistics.

In the following, we study the spin relaxation under identical
strengths of the Dresselhaus and the Rashba couplings and with the cubic term of
the Dresselhaus term included.
We first show the temporal evolution of the spin
polarization along all directions, {\em i.e.}, $\mathbf{\hat n}_1$,
$\mathbf{\hat n}_2$ and $\mathbf{\hat{z}}$ in these two
statistics. Then we investigate the SRT along the direction
$\mathbf{\hat n}_1$  with the impurity scattering and the
electron-LO phonon scattering included to compare the effect of different statics
to the SRT. Finally we give the SRT with all the
scattering including at different temperatures, electron and
impurity densities  and well widths with the collinear statistics.

In the numerical calculation, we only need to consider electron-longitudinal optical (LO)
phonon scattering for the electron-phonon scattering as
we focus on the high temperature regime ($T>120$\ K). The matrix
element of the electron-LO phonon interaction is
$|g_{\mathbf{q}q_z,LO}|^2 =
\{\alpha\Omega_{\mbox{LO}}^{3/2}/[\sqrt{2\mu}(q^2+q_z^2)]\}|I(iq_z)|^2$
with $\alpha=e^2\sqrt{\mu/(2\Omega_{\mbox{LO}})}
(\kappa^{-1}_{\infty}-\kappa_0^{-1})$. $\kappa_{\infty}$ is the
optical dielectric constant and
$\Omega_{\mathbf{q}q_z}=\Omega_{LO}$ is the LO-phonon
frequency. The material parameters of GaAs are listed
as following:\cite{made}
$m^{\ast}=0.067 m_0$ with $m_0$ the free electron mass,
$\kappa_{\infty}=10.8$, $\kappa_0=12.9$ and $\Omega_{LO}=35.4$\ meV,
The strengths of the Dresselhaus SOC and Rashba SOC are $\gamma=25$\
meV \r{A}$^3$ and $\alpha=\gamma (\frac{\pi}{a})^2$  with $a=5$\
nm unless otherwise specified.\cite{Loss}

\begin{figure}[h]
\psfig{file=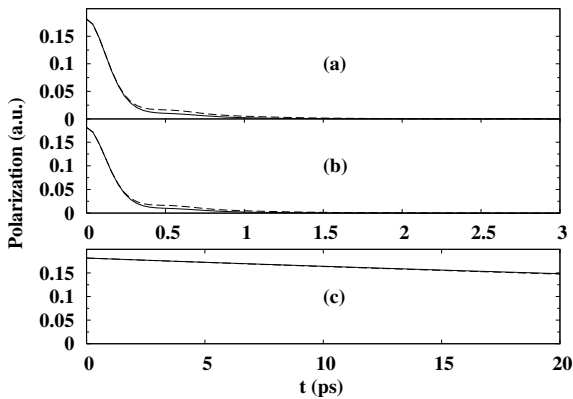,width=8.cm}
\caption{The temporal evolutions of spin polarizations along
(a): ${\bf {\hat z}}$; (b): ${\bf{\hat n}}_2$; (c): ${\bf{\hat n}}_1$
in the two statistics with
  $N_e=4\times10^{11}$\ /cm$^2$ and $N_i=0.5N_e$ at $T=200$\ K.
Solid curves: in the collinear statistics;
  Dashed ones: in the helix statistics.
}
\label{fig:temporal}
\end{figure}

\subsection{Temporal evolutions of spin polarization in collinear and helix statistics}
\label{sectiona}
In Fig. \ref{fig:temporal}, we plot the
temporal evolutions of the spin polarization along the three
directions: ${\bf{\hat z}}$ [Fig.\ \ref{fig:temporal}(a)],
${\bf {\hat n}}_2$ [Fig.\ \ref{fig:temporal}(b)] and ${\bf {\hat n}}_1$ [Fig.\
\ref{fig:temporal}(c)] with the
electron-LO phonon and the electron-impurity scattering
included at temperature $T = 200$\ K in the system with identical
strengths of the Rashba and Dresselhaus coupling. The impurity density
$N_i$ is $0.5N_e$. The evolutions within these two statistics are
plotted in the same figures for comparison.

It is seen from the figure that with the cubic Dresselhaus term included,
the SRT along ${\bf{\hat n}}_1$ becomes finite (several hundred picoseconds),
 but still much longer than
the other two perpendicular directions
${\bf{\hat z}}$ and ${\bf{\hat n}}_2$ (less than 1\ ps). This is because of the
strong anisotropy from the EMF $\mathbf{h}(\mathbf{k})=h_1(\mathbf{k})
\mathbf{\hat{n}_1}+h_2(\mathbf{k})\mathbf{\hat{n}_2}$
  with $h_1(\mathbf{k})=\sqrt{2}\gamma(k_y-k_x)[(\frac{\pi}{a})^2
  + \frac{k_xk_y}{2}]$ and $h_2(\mathbf{k})=\gamma
  k_xk_y\frac{k_y+k_x}{\sqrt{2}}$. As the spin flip is determined by the
  component of the EMF which is perpendicular
to the spin polarization, therefore different direction of the spin polarization
experiences different EMF: for electron spin along the direction
${\bf{\hat z}}$, it feels the total EMF $\mathbf{h}(\mathbf{k})$;
for the one along ${\bf {\hat n}}_{2(1)}$, it feels $h_{1(2)}(\mathbf{k})$.
For narrow QW's, the linear term is
much larger than the cubic terms and it appears only in $h_1({\bf k})$.
Therefore the spin along the direction ${\bf {\hat n}}_{1}$ feels only the
cubic terms and hence experiences a much weaker spin relaxation; whereas the
spin along the other two directions feels the linear term and
shows much shorter and similar SRT's.

It is noted that the difference of the time evolutions of the
spin polarizations between the helix and collinear statistics is
marginal in all directions:
The evolutions of spin polarization
along ${\bf{\hat n}}_1$ are almost identical while those along ${\bf{\hat n}}_2$ and
${\bf{\hat z}}$ show marginal differences after the initial fast spin relaxation.
This is because the spin-orbit couplings along the latter two directions are much
stronger than the one along ${\bf{\hat n}}_1$.
Nevertheless, it is shown that even for the very strong spin-orbit coupling
strengths used in our calculation, the
difference of the time evolution of the spin
polarization between the helix and the collinear statistics
is too marginal to affect the value of the SRT effectively. Therefore, for the spin
polarized system, the statistics used in the scattering do not affect the SRT too much.
We will further show this in the next subsection focusing on the spin relaxation
along ${\bf{\hat n}}_1$ as we are interested in the
longest SRT along this direction.

\subsection{Comparison of the SRT along direction (1,1,0)
  with helix and collinear statistics}
\label{sectionb}

\begin{figure}[h]
\psfig{file=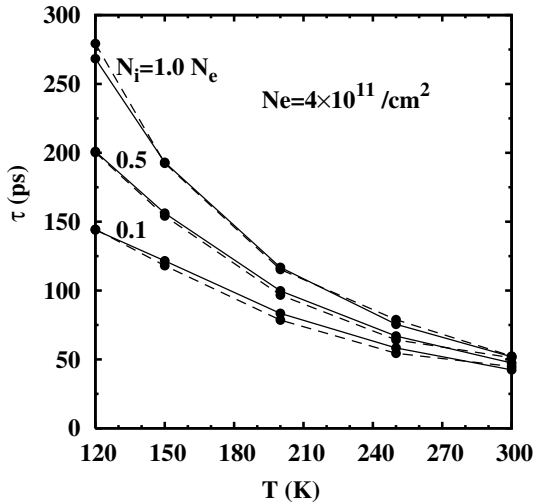,width=7.cm}
\caption{Spin relaxation time {\em vs.} the temperature for different impurity
  densities $N_i=0.1N_e$, $N_i=0.5N_e$ and $N_i=1.0N_e$ in the two
statistics with the electron density
  $N_e=4\times10^{11}$\ cm$^{-2}$. Solid curve: with the helix
  statistics; Dashed curve: with the collinear statistics.}
\label{fig:compare}
\end{figure}

In this subsection we compare the SRT's along the
special direction ${\bf{\hat n}}_1$ in
the helix and collinear statistics with the electron-non-magnetic impurity
and the electron-LO phonon scattering included.
In Fig.\ \ref{fig:compare} we plot the SRT as a function of the temperature
for different impurity densities $N_i=0.1N_e$, $0.5N_e$ and $1.0N_e$,
with the electron density $N_e=4\times10^{11}$\ cm$^{-2}$. The dots
with solid (dashed) curves are the results with the helix (collinear) statistics.
It is seen from the figure that in both statistics the SRT's along
${\bf{\hat n}}_1$
 decrease with increase of the temperature or
with decrease of the impurity density. These results are consistent  with
our previous calculations,\cite{wu2,wu21,strain} as the inhomogeneous broadening
induced by the EMF here depends cubically on the momentum ${\bf k}$. We have
shown in our previous work\cite{wu2,strain} that in GaAs QW's when the cubic term of the
Dresselhaus is dominant, the SRT decreases with the temperature.

From the figure one further finds that the SRT along ${\bf{\hat n}}_1$ in
the two statistics are almost the same. This is consistent with the
result in the previous subsection. Thus it is suitable to  study the SRT with
one specific statistics, {\em eg.}, in the collinear statistics as the scattering
can be treated more precisely and efficiently.

\subsection{SRT along the direction (1,1,0) with the electron-electron
  scattering included in the collinear statistics}
\label{sectionc}

\begin{figure}[h]
  \centering
  \psfig{file=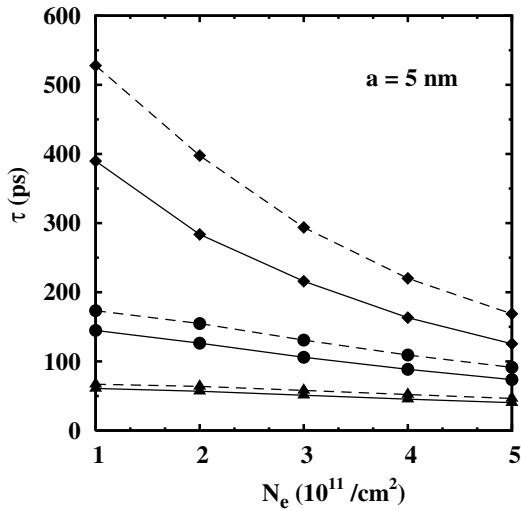,width=7cm}
  \caption{Spin relaxation time {\em vs.} the electron density for
    different impurity densities $N_i=0$ (solid curves) and $0.5N_e$
    (dashed curves) at  $T=120$\  K ($\blacklozenge$),
200\ K ($\bullet$) and 300\ K ($\blacktriangle$) with all the
    scattering included. The well width $a=5$\ nm.}
  \label{fig:NiT-C}
\end{figure}

In the previous two subsections, we draw the conclusion that the
SRT in the two statistics are almost the same. Although we did not
include the electron-electron Coulomb scattering above as it is extremely
difficult to compute the coulomb scattering quantitatively in the helix statistics
numerically  to meaningfully compare the SRT from the
collinear statistics for the reason given in the previous section, we still
believe inclusion of the Coulomb scattering would not bring a vast change
to the SRT from the two statistics, especially for the polarization along
${\bf{\hat n}}_1$ where the spin-orbit coupling is weak.
Moreover, recent studies of spin
kinetics have shown that the Coulomb scattering plays a crucial role
in spin R/D.\cite{wu1, wu, wu2,wu21,ivan}
Therefore in this subsection we investigate the properties of the SRT along ${\bf{\hat n}}_1$
 in the collinear statistics with the Coulomb scattering explicitly included.

In Fig.\ \ref{fig:NiT-C} we plot the
SRT versus the electron density with and without
impurities at different temperatures T=120\ K ($\blacklozenge$), 200\
K($\bullet$) and 300\ K  ($\blacktriangle$).
The solid curves are for
case without impurity and the dashed ones are for the case with impurity
($N_i=0.5N_e$). We find the impurity and
the temperature dependence of the SRT with the electron-electron Coulomb
scattering included is all the same with that in
Sec.\ \ref{sectionb} without the Coulomb scattering.
By comparing the values at the same conditions in Fig.\
\ref{fig:compare}, it is easy to find that the SRT with the
electron-electron scattering is larger, which is consistent
with the previous results.\cite{wu3,wu2,wu21}. It is also seen from the
figure that the SRT decreases with the increase of the electron
density and the rate of decrease is slower with increase
of the temperature. It can be understood from the view of the
inhomogeneous broadening:\cite{wumany}  For higher electron density,
electrons tend to populate to higher momentum space. Therefore the
inhomogeneous broadening induced by the EMF becomes larger which results in a shorter
SRT. Moreover, when the temperature is high enough, electrons prefer
to distribute in high momentum space even at low electron densities.
Therefore the inhomogeneous broadening becomes less
sensitive to the change of the electron density. This explains why the
SRT becomes insensitive to the electron density at at high temperatures.

\begin{figure}[h]
  \centering
  \psfig{file=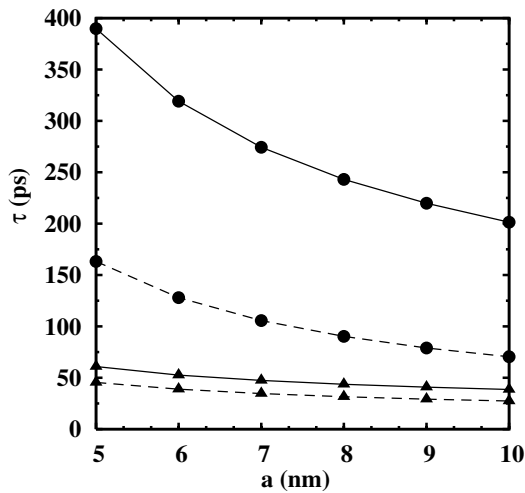,width=7cm}
  \caption{Spin relaxation time {\em vs}. the width of the QW
    for electron densities $N_e=1\times10^{11}$\ cm$^{-2}$
(solid curves) and $4\times10^{11}$\ cm$^{-2}$ (dashed curves)
at different temperature 120\ K (curves with $\bullet$) and 300\
    K (curves with $\blacktriangle$). The impurity density $N_i=0$.}
  \label{fig:Ne-a}
\end{figure}

Now we turn to the well width dependence of the SRT.
It is noted that when the well width is changed, the relation
  $\beta=\gamma(\frac{\pi}{a})^2$ is maintained by changing the
  external gate voltage.
In Fig.\  \ref{fig:Ne-a} we plot the SRT versus the width of the
QW for different electron densities
$N_e=1\times10^{11}$\ cm$^{-2}$ and $4\times10^{11}$\ cm$^{-2}$
at $T=120$\ K and 300\ K.
The impurity is taken to be zero in the calculation.
{\em Contrary} to the previous results in (001) QW's with only the
Dresselhaus spin-orbit coupling,\cite{wu2,wu21}
here one finds that the SRT {\em decreases} with the increase of the well width.
This is because in the present case with identical strengths
of the Rashba and Dresselhaus spin-orbit coupling,
 the component of the EMF which can flip electron spin along ${\bf {\hat n}}_1$-direction
is only the cubic term of the Dresselhaus coupling, which is independent
on the well width. Nevertheless,
the linear EMF along the direction  ${\bf {\hat n}}_1$ is
proportional to $a^{-2}$ which becomes larger
with decrease of the well width.
For large linear EMF along ${\bf {\hat n}}_1$, electron spins
prefer to line along ${\bf {\hat n}}_1$. Therefore
spin flipping along ${\bf {\hat n}}_1$ is suppressed and thus the inhomogeneous broadening
which leads to the spin relaxation along ${\bf {\hat n}}_1$ is suppressed.
Consequently the SRT increases with decrease of the well width.

\section{Conclusions}
In conclusion, we study the spin relaxation under the identical
Rashba and Dresselhaus coupling strengths in (001) GaAs QW by the
kinetic spin Bloch equations with different statistics: the collinear
 and the helix statistics. We show that the SRT's calculated from both the helix
statistics and collinear statistics differ marginally in all directions.
This is very important for the study of the spin R/D as most spin R/D calculations
are performed with the collinear statistics and the numerical evaluation
of the scattering within the
collinear statistics is much more accurate and faster.

We show that when the cubic term of the Dresselhaus spin-orbit coupling is ignored,
the SRT's along (1,1,0) direction in both statistics are infinite. However, with the
inclusion of the cubic term, the spin relaxation along this direction becomes
finite, although still much longer than that along the other two perpendicular directions.
We study impurity density, temperature, electron density
and the well width dependence  of the SRT along this direction.
We find the SRT
increases with increase of the impurity density and  decreases with
increase of the temperature or the electron density. These
properties are all the same with case of (001) QW with only the Dresselhaus spin-orbit
coupling. However, the SRT decreases
with increase of the well width, which is {\it contrary} to
the previous results. All these behaviors are explained following our
previous theories.

\begin{acknowledgments}
This work was supported by the Natural Science Foundation of China
under Grant Nos. 90303012 and 10574120, the Natural Science
Foundation of Anhui Province under Grant No. 050460203, the
Innovation Project of Chinese Academy of Sciences and SRFDP. One
of the authors (M.W.W.) was also partially supported by SORST
program from JST. He would like to thank Vladimir Privman at
Clarkson University and Makoto Kuwata-Gonokami at Tokyo University
for hospitality where this work was finalized. J.L.C. would like
to thank Mr. L. Jiang for running some of the codes for him.
\end{acknowledgments}

\appendix
\section{The  kinetic spin Bloch equations in the helix spin space}
\label{helix_formula}

In the helix spin space, the physics
meaning of each matrix elements $\rho^{h}_{{\bf k},\eta,\eta^\prime}
=\langle\eta|\rho_{\bf k}|\eta^\prime\rangle$ is not well defined as the helix
spin state is a mixture of spin-up and -down states and is ${\bf k}$
dependent [Eq.\ (\ref{helix})].  Here the superscript $h$ denote the
quantity in the helix spin space. The density matrices and the kinetic
spin Bloch equations in the helix spin spaces can be transformed from
that in the collinear spin space by a unitary
  transformation:
  $\rho^{h}_{\mathbf{k}}=U_{\mathbf{k}}^{\dag}\rho_{\mathbf{k}}
U_{\mathbf{k}}$, with
\begin{equation}
U_{\mathbf{k}}=\frac{1}{\sqrt{2}}\left(\begin{array}{cc}1&1\\ \frac{{\tilde h}(\mathbf{k})}
      {|{\bf h}(\mathbf{k})|} & -\frac{{\tilde h}(\mathbf{k})}
      {|{\bf h}(\mathbf{k})|}\end{array}\right)\ .
\end{equation}
The coherent terms are
\begin{eqnarray}
 \dot{f}_{\mathbf{k},+}^{h}|_{\mathtt{coh}} &=&-\dot{f}_{\mathbf{k},-}^{h}|_{\mathtt{coh}}=
2\mbox{Im}[P^{\ast h}_{\mathbf{k}}\rho_{\mathbf{k},+-}^{h}]\ ,\\
\dot{\rho}_{\mathbf{k},+-}^{h}|_{\mathtt{coh}}
&=& -i
P_{\mathbf{k}}^{h}(f_{\mathbf{k},+}^{h}-f_{\mathbf{k},-}^{h}) - i
\Delta_{\mathbf{k}}^{h}\rho_{\mathbf{k},+-}^{h}\ ,
\end{eqnarray}
in which
\begin{widetext}
\begin{eqnarray}
P_{\mathbf{k}}^h&=&-\sum_{\mathbf{q}}V_{\mathbf{q}}
\{\mbox{Re}[\rho_{\mathbf{k-q}}^h]+i\mbox{Re}[h^{\prime}
(\mathbf{k},\mathbf{q})]\mbox{Im}[\rho_{\mathbf{k-q},+-}^h]+i
\mbox{Im}[h^{\prime}(\mathbf{k},\mathbf{q})]
(f_{\mathbf{k-q},+}^h-f_{\mathbf{k-q},-}^h)/2\}\ ,\\
\Delta_{\mathbf{k}}^h&=&2|\mathbf{h}(\mathbf{k})|
-\sum_{\mathbf{q}}V_{\mathbf{q}}\{\mbox{Re}[h^{\prime}(\mathbf{k},\mathbf{q})]
(f_{\mathbf{k-q},+}^h-f_{\mathbf{k-q},-}^h)-\mbox{Im}[h^{\prime}
(\mathbf{k},\mathbf{q})]\mbox{Im}[\rho_{\mathbf{k-q},+-}^h]\}\ .
\end{eqnarray}
In these equations  $h^{\prime}(\mathbf{k},\mathbf{q})
={\tilde h}^{\ast}(\mathbf{k}){{\tilde h}(\mathbf{k-q})}/
[{|{\tilde h}(\mathbf{k})|}
      {|{\tilde h}(\mathbf{k-q})|}]$.
The scattering terms are
\begin{eqnarray}
  \dot{\rho^h}_{\mathbf{k}}|_{\mathtt{scatt}} &=&\bigg\{
  \pi N_i\sum_{\mathbf{q}}|U_{\mathbf{q}}|^2
  \sum_{\xi_1,\xi_2}
  \delta(\varepsilon_{\mathbf{k}-\mathbf{q},\xi_1}-\varepsilon_{\mathbf{k},\xi_2})
  S_{\mathbf{k},\mathbf{k-q}}T^h_{\mathbf{k}-\mathbf{q},
    \xi_1}(S_{\mathbf{k-q},\mathbf{k}}\rho^h_{\mathbf{k}}-\rho^h_{\mathbf{k}-\mathbf{q}}
    S_{\mathbf{k-q},\mathbf{k}})T^h_{\mathbf{k},\xi_2} \nonumber
\\  &+&
  \pi\sum_{\mathbf{q}q_z\lambda}|g_{\mathbf{q}q_z\lambda}|^2
  \sum_{\xi_1,\xi_2}S_{\mathbf{k},\mathbf{k-q}}T^h_{\mathbf{k}-\mathbf{q},\xi_1}
  \nonumber\\
&&\mbox{}\times\big\{\delta(\varepsilon_{\mathbf{k}-\mathbf{q},\xi_1}-\varepsilon_{\mathbf{k},\xi_2}
  + \Omega_{\mathbf{q}q_z\lambda})
  [(N_{\mathbf{q}q_z\lambda}+1)(1-\rho^h_{\mathbf{k}-\mathbf{q}})S_{\mathbf{k-q},\mathbf{k}}
  \rho^h_{\mathbf{k}}\nonumber\\
&&\mbox{}-
N_{\mathbf{q}q_z\lambda}\rho^h_{\mathbf{k}-\mathbf{q}}S_{\mathbf{k-q},\mathbf{k}}
(1-\rho^h_{\mathbf{k}})]\nonumber\\
  &&\mbox{}+\ \delta(\varepsilon_{\mathbf{k}-\mathbf{q},\xi_1}
  -\varepsilon_{\mathbf{k},\xi_2}-
  \Omega_{\mathbf{q}q_z\lambda})[N_{\mathbf{q}q_z\lambda}(1-\rho^h_{\mathbf{k}-\mathbf{q}})
  S_{\mathbf{k-q},\mathbf{k}}
   \rho^h_{\mathbf{k}}\nonumber\\
&&\mbox{}-(N_{\mathbf{q}q_z\lambda}+1)\rho^h_{\mathbf{k}-\mathbf{q}}S_{\mathbf{k-q},\mathbf{k}}
(1-\rho^h_{\mathbf{k}})]\big\}T^h_{\mathbf{k},\xi_2}\nonumber
\\
 &+&
  \pi\sum_{\mathbf{q}\mathbf{k}^{\prime}}V_{\mathbf{q}}^2\sum_{\xi_1, \xi_2, \xi_3,
    \xi_4}\delta(\varepsilon_{\mathbf{k}^{\prime},\xi_3}
-\varepsilon_{\mathbf{k}^{\prime}-\mathbf{q},\xi_4}+\varepsilon_{\mathbf{k}
-\mathbf{q},\xi_1}-\varepsilon_{\mathbf{k},\xi_2})S_{\mathbf{k},\mathbf{k-q}}
T^h_{\mathbf{k}-\mathbf{q},\xi_1}\nonumber
\\
 &&\times\big\{\mbox{Tr}[T^h_{\mathbf{k}^{\prime},\xi_3}(1-\rho^h_{\mathbf{k}^{\prime}})
 S_{\mathbf{k}^{\prime},\mathbf{k}^{\prime}-\mathbf{q}}
 \rho^h_{\mathbf{k}^{\prime}-\mathbf{q}}T^h_{\mathbf{k}^{\prime}-\mathbf{q},\xi_4}S_{\mathbf{k}^{\prime}
 -\mathbf{q},\mathbf{k}^{\prime}}]
 (1-\rho^h_{\mathbf{k}-\mathbf{q}})S_{\mathbf{k-q},\mathbf{k}}\rho^h_{\mathbf{k}}\nonumber\\
&& -
  \mbox{Tr}[T^h_{\mathbf{k}^{\prime},\xi_3}\rho^h_{\mathbf{k}^{\prime}}S_{\mathbf{k}^{\prime},
  \mathbf{k}^{\prime}-\mathbf{q}}
(1-\rho^h_{\mathbf{k}^{\prime}-\mathbf{q}})T^h_{\mathbf{k}^{\prime}-\mathbf{q},\xi_4}
S_{\mathbf{k}^{\prime}-\mathbf{q},\mathbf{k}^{\prime}}]
\rho^h_{\mathbf{k}-\mathbf{q}}S_{\mathbf{k-q},\mathbf{k}}(1-\rho^h_{\mathbf{k}})\big\}
T^h_{\mathbf{k},\xi_2}\bigg\}
\nonumber
\\
 &+&\Big\{\cdots \Big\}^{\dagger}
 \label{scat:helix}
\end{eqnarray}
\end{widetext}
with
$S_{\mathbf{k},\mathbf{k-q}}=U_{\mathbf{k}}^{\dag}U_{\mathbf{k-q}}=\frac{1}{2}[
(1+h^{\prime}(\mathbf{k},\mathbf{q}))+
(1-h^{\prime}(\mathbf{k},\mathbf{q}))\sigma_x]
$
and
$T^h_{\mathbf{k},\xi}=U_{\mathbf{k}}^{\dag}T_{\mathbf{k},\xi}U_{\mathbf{k}}$.
In the helix spin space,
 $T^{h}_{\mathbf{k},\xi}=T^{h}_{\mathbf{k},\sigma}=1/2$ for the
collinear statistics and $T^h_{\mathbf{k},\xi}=T_{{\mathbf k},\eta}^h
=\frac{1}{2}[1+\eta\sigma_z]$ for the helix statistics.
%\newpage

\end{document}